\documentclass[12pt]{article}
\usepackage{graphicx, mdframed}
\usepackage{mathptmx}     
\usepackage[T1]{fontenc}

\usepackage{eulervm}
\usepackage{booktabs}

\usepackage{amsmath}
\usepackage{amssymb}
\usepackage{amsthm}

\usepackage{graphicx}
\usepackage{subcaption}

\begin{document}

\title{Elementary Bitcoin economics: from production and transaction demand to values}

\author{Misha Perepelitsa}

\maketitle

\begin{abstract}

In this paper we give an elementary analysis of economics of Bitcoin that combines the transaction demand by the consumers and the supply of hashrate  by miners. We argue that the decreasing block reward  will have a lesser effect on the exchange rate (price) of Bitcoin and thus the network will be transitioning to a regime where transaction fees will play a bigger part of miners' revenue. We consider a simple model where consumers demand bitcoins for transactions, but not for hoarding bitcoins, and we analyze market equilibrium where the demand is matched with the hashrate supplied by miners.
 
Our main conclusion is that the exchange rate of Bitcoin cannot be determined from the market equilibrium and so our arguments support the hypothesis that Bitcoin price has no economic fundamentals and is free to fluctuate according the present demand for hoarding and speculation.
We point out that increasing fees  bear the risk of Bitcoin being outcompeted by its main rival Ethereum, and that decreasing revenues to miners  depreciate the perception of Bitcoin as a medium for store value (hoarding demand) which will have effect its exchange rate.
\end{abstract}

\section{Introduction} 
Bitcoin is an exciting socio-economic phenomenon. With all its fabulous growth and recent breathtaking plunges, it is a powerful motivator for the participants, investors and  miners alike, while its philosophy is a source of inspiration for many progressive technocrats.
 
Despite these evident economic inefficiencies of the Bitcoin market due to speculation, herding, and social media effects (Garcia et al.  (2014), Cheah and Fry (2015), Goczek  and  Skliarov (2019)), it is  clear that in these 13 years since its inception, the Bitcoin network proved to be stable and robust and it has delivered what it claimed: the distributed ledger with clever consensus mechanism that prevents accounting fraud.

The functionality of the network relies on the working of large number of miners, who  provide the computational power to process transactions and to build the blockchain.  As the basic computation operation is to generate a hash value and to check that it has a predetermined property, it is natural to measure miners' output in hashes per second, or multiple of that like terahashes. Producing hasrate involves costs, that include equipment, energy and labor. Thus, a properly working network must incentives miners for their work, which, in the case of Bitcoin, comes from transaction fees and block reward.  On the consumer side, people need bitcoins to carry transaction on the network. Hoarding bitcoins is a popular activity which also adds to the demand for bitcoins. All this creates a classical economic paradigm of production, consumption and value.

In this paper we use elementary tools of economics to address the basic questions of Bitcoin economy: how the price of Bitcoin determines the hashrate supplied to the network, what the competition between miners for a bigger share of hashrate does to the supply and the revenue of miners,   
what the optimal (equilibrium) outcome for the transaction demand and supply of hashrate is.

Our main motivation for this analysis is to gain some clues on the future of Bitcoin. Thus, we will discuss such things as the halving of the block rewards, a possibility of increasing transaction fees and how all this will reflect on the price of Bitcoin and its dominance among other cryptocurrencies.

 The marginal cost of production formula that we use in section \ref{sec:supply} first appeared in Garcia et al. (2013).  It was further developed by Hayes (2015) who introduced the cost of production formula for Bitcoin price. Statistical tests were performed by Hayes (2016, 2019) to compare the price predicted by the cost of production formula with the observed Bitcoin price for the period 2013-2018,
showing good agreement between two time series. A similar conclusion was drawn in Abbatemaraco et al. (2018). Contrary to that, a statistical study by Baldan and Zen (2020) found that Bitcoin historical prices are not connected with the price derived from the cost of production model, and it was suggested that the discrepancy between two studies could be due to the different time frames. 

The point of view we take in this paper is that the cost of production formula {\it alone} is not sufficient to determine the price. It only determines the supply of hashrate (under competition) for given price. It can be used to find price only if we know that  the market has reached an equilibrium where demand and supply balance each other.   This can also explain the differences in the empirical studies mentioned above. In different time frames, the market can be closer or further away from an equilibrium, because of an appearance of transient bubbles or simply because  the equilibrium itself changes due to changing market environment.

Our model is based on the assumption, discussed in some details in section \ref{sec:halving}, that with over 90\% of bitcoins already in circulation, the remaining bitcoins will have lesser effect on the price of Bitcoin and the block reward to miners will have diminishing contribution to the miners' revenue. The more bitcoins are mined, the less effect the last bitcoins will have on the price. With that, we consider (section \ref{sec:demand}) an economy where all revenue of miners comes from transaction fees. Assuming elasticity of transaction demand, we show that there is a market equilibrium and such equilibrium does not involve the exchange rate (price) of Bitcoin. Instead, it determines the price of the hashrate. As the main function of Bitcoin is to transact money, our argument adds evidence in support of a hypothesis that Bitcoin price is not based on economic fundamentals, a claim that was made many times before, for example, Yermack (2013), Hanley (2013),  Cheah and Fry (2015). 

We argue further that the transition to a regime where fees dominate block reward carries with it uncertainty about the future of Bitcoin, due the increase in transaction fees and negative effect of declining miners' revenue on price of Bitcoin. The latter might undermine the popular perception of Bitcoin as a store of value.

People tend to underestimate the risk of bitcoin being out-competed by other cryptocurrencies and over-estimate the store value of Bitcoin. In fact, the idea that bitcoin is good for saving is so popular that some experts even state that it is a function of Bitcoin to store value.  In reality, what is true is that Bitcoin is good for storing only the numerical value of your bitcoins but nothing beyond that.

If you're investing into Bitcoin, it should be clear that is not some intrinsic value that you're purchasing. You're betting your money that from {\it now} to the point in the {\it future} when you're going to sell bitcoins, the rate of investment into Bitcoin will keep growing at least as fast as the inflation. Moreover, you need to be lucky enough so that when you will be selling your bitcoins, others are still buying into it.

An argument is often put forward that the store value somehow comes from the limited quantity of bitcoins in existence, like the value of Rembrandt paintings or diamonds. That is, from things that are rare and that have come to be markers of social status or financial success. 

Yet, viewed as a part of the whole crypto-ecosystem it is unlimited. Bitcoin is not unique in its functions and it has no value besides these functions. The essence of the bitcoin is the blockchain algorithm that is publicly available, is implemented in all existing crypto-currencies and can be effortlessly duplicated to create yet another.  The perceived uniqueness of Bitcoin is that it is the first network of its kind and at the moment has largest capitalization. However, it is not rare in a sense of paintings of famous artists or diamonds.  If there is another cryptocurrency that offers more for less for both consumers and miners the dominance of Bitcoin will be broken and with it, its value.

Perhaps it is true that crypto is the future of money.  At present, however, speculation and extreme volatiliy are main obstalce for wider  acceptence of Bitcioins and other cryptocyrrencies.  This calls for the technological innovations of the kind that will address the economic issues of blockchain algorithms.

\begin{table}
\centering
\begin{tabular}{@{} *5l @{}}    \toprule
\emph{label} & \emph{description} & units   \\\midrule
X    & exchange rate (price) of the bitcoin  & USD/BTC   \\ 
H   & network hashrate & tH/s \\ 
p   & price of electricity & USD/kWh\\
F   & network daily transaction fees & USD/day  \\
BR & network daily block reward & BTC/day  \\
 \bottomrule
 \hline
\end{tabular}
\caption{Notation used in the paper. tH stands for tera-hashes.}
\label{tab:notation}
\end{table}

\section{Supply of hashrate}
\label{sec:supply}
\begin{figure}[ht]
\begin{minipage}[b]{0.45\linewidth}
\centering
\includegraphics[width=\textwidth]{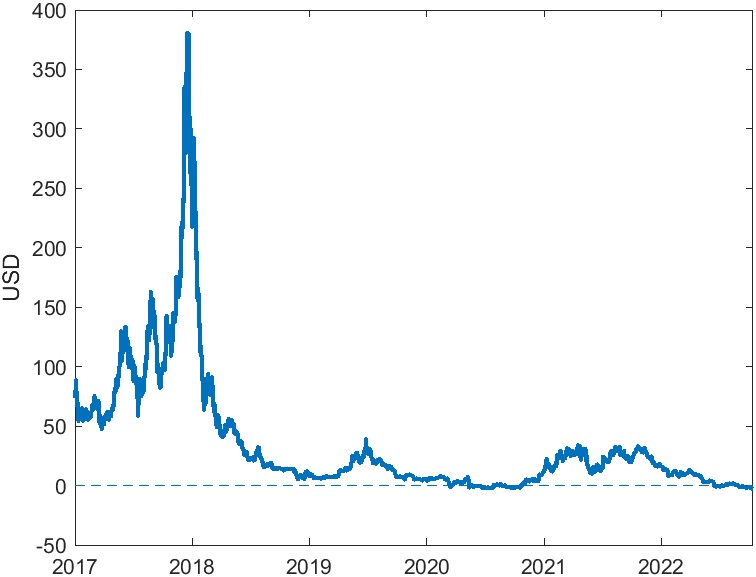}
\end{minipage}
\hspace{0.5cm}
\begin{minipage}[b]{0.45\linewidth}
\centering
\includegraphics[width=\textwidth]{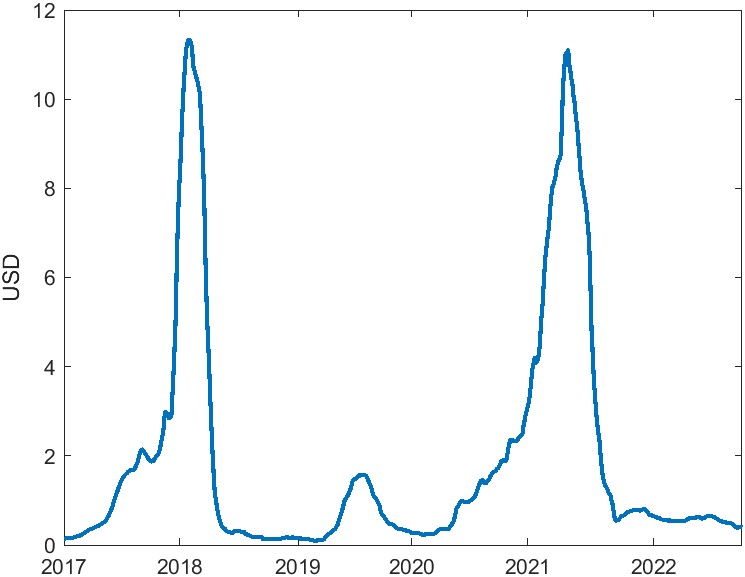}
\end{minipage}
\caption{(Left) Daily profit of a 100tH/s miner with energy consumption $\theta=\ 3$kWh and the price of electricity $p=0.15$ USD/kWh. The plot was prepared using data from https://coinmetrics.io/  (Right) Median transaction fee on the Bticoin network averaged over period of 200 days (from https://coinmetrics.io/)}
\label{fig:1}
\end{figure}

Table \ref{tab:notation} lists the variables that we are use in the paper. We start with the analysis of marginal profit for operating computational nodes of Bitcoin. One can compute a marginal profit of a miner for adding, let’s say, a 100tH/s mining ASIC with energy consumption (power) $\theta=\ 3$kW and the price of electricity $p=0.15$USD/kWh.  Without taking into consideration the cost of the equipment and labor we get the marginal profit (MP) as the difference between the marginal revenue (MR) and the marginal cost (MC):
\[
MP\ =\ MR-MC=\left( F+X\cdot BR\right)\frac{100}{H}-24\theta p
\]
This is a marginal profitability and it is openly available on Internet. For example, as of October 15, 2022, using one-week averages for X, F and BR we compute that the margin profit equals -3 USD. The profitability obviously is affected by the price of electricity, p, which greatly varies geographically. Here, it is quoted as an average price of electricity in USA. Historically, mining of bitcoins was a profitable enterprise with large margins. Figure \ref{fig:1} (left plot) shows the profitability (USD per day) for operating a 100tH/s miner for the last three years, with the parameters used in the above calculations. The profitability slumped to the region around zero in the second half of 2022, which is attributed to the influx of miners from Ethereum network to Bitcoin due to the change of the Ethereum’s mining algorithm. 

In the above formula we use units of BTC for block reward BR because it is determined by the network. Thus the  revenue from block reward is $X\cdot BR.$ Whereas we use  US dollars for the transaction fees because transaction fees, expressed in USD, do not exhibit any significant temporal trends  see Figure \ref{fig:1}. In that figure, three anomalous  peaks correspond to times of faster than exponential growth of Bitcoin.

Miners use formula (1) to switch on and off the mining equipment or invest in a new one. In  a competitive environment of many independent miners, one can expect 
 \begin{equation}
 \label{eq:1}
 F+X\cdot BR=\frac{24\theta pH}{100}.
 \end{equation}
The formula like \eqref{eq:1} was used in Garcia et al. (2013) and in Hayes (2015, 2016, 2019) for determining exchange rate (price) $X.$ That however, implicitly, assumes that the price corresponds to the market equilibrium between supply and demand. Without this assumption, formula \eqref{eq:1} only determines the supply of hashrate for a given price $X$ in a competitive market.

In the competitive market each miner is trying to out-compete others by increasing hashrate output. This however does not increase the whole network revenue, because of the automatic adjustment of difficulty. Thus, the outcome of competition is the increase of hashrate (and energy costs) and zero increase in the revenue.  This process also shows is the main inefficiency of the Bitcoin network that has been pointed out by many authors. Miners could produce, for example, half as much hashrate with no effect on the functionality of the network, consumers and miners' revenue.


If there are just few firms controlling most of the hashrate, the marginal profit analysis is slightly different. Let $k_1,..,k_n$ be the proportions of hashrate produced by $n$ firms. For firm i,  the daily profit is
\[ 
P_i=k_i (  F+X\cdot BR-24\theta pH/100).
\]
Let’s consider the effect of adding one 100Th/s computational unit (ASIC) to the production line of firm 1. The change in the profit for this firm:
\begin{equation}
\label{eq:d1}
\Delta P_1 = \frac{100(1-k_1)}{H+100} (F+X\cdot BR)-24 \theta p,
\end{equation}
and for all other  firms,
\begin{equation}
\label{eq:d2}
\Delta P_i=-\frac{100k_i}{H+100}(F+X\cdot BR)<0,\quad i>1.
\end{equation}
Note that the daily block reward BR does not change on average with variations of hashrate H, because Bitcoin algorithm automatically adjusts the difficulty of the computational task.
Expression in \eqref{eq:d2} is negative, since the costs for firm $i\, (i>1)$ didn’t change but the proportion of the hashrate it produces decreased. Thus if $\Delta P_1>0,$ firm 1 adds a unit to production, but other firms will follow the trend, leading to certain redistribution of production proportions $k_1,..,k_n.$ Assuming, for simplicity, that $k_1=..=k_n,$ we can get zero marginal profit condition as 
\[
\frac{100\left(1-1/n\right)}{H+100}\left(F+X\cdot BR\right)=24\theta p.
\]
Or, since H is much larger than 100,
\begin{equation}
\label{eq:firms}
\frac{100\left(1-1/n\right)}{H}\left(F+X\cdot BR\right)=24\theta p.
\end{equation}
At such level of H, firms have non-zero profit:
\[
P_i=\frac{F+X\cdot BR}{n^2},\quad i=1..n.
\]

\subsection{Effect of varying energy costs}

One must be careful in interpreting equation \eqref{eq:1} or \eqref{eq:firms}. For example, it is wrong to claim that rising electricity costs do necessarily lead to the increase of exchange rate $X$ ``to compensate miners for the lost profit.'' The claim is of course misleading, implying that miners are guaranteed a certain level of income. If the electricity costs rise, some miners (inefficient ones or operating in the countries with highest electricity) will leave the market, increasing the share of the revenue of the remaining miners. 

Moreover,  the network will decrease the computational difficulty (making sure that the same work is done) which will  decrease the energy costs of the remaining miners. Thus, the hashrate will drop but the right hand-side of equation \eqref{eq:1} or \eqref{eq:firms} will remain the same. Likewise, when electricity costs are down, the hashrate goes up, with new miners joining the network.

\subsection{Bitcoin halving events}
\label{sec:halving}
Every 4 years the block reward in BTC is halved by the bitcoin algorithm, reducing the revenue of miners by the same order, because the transaction fees make up a small fraction of the revenue.

Sometimes it is claimed that after a halving event the price X must double  because the supply of new bitcoins is cut in half. This statement is based on false premises about the nature of the demand for bitcoins. It is unlikely that a person demands a certain amount of bitcoins per unit of time, without any regard how much she already has, as with the demand for perishable necessities. 

It is not impossible for the demand for bitcoins to be per unit of time demand during some periods of time.  For example, the influx of new market participants can create this effect, and it might have been the reason why Bitcoin price increased after halving events in the past.  This, however long it can be, is a transient phenomenon, and the consideration such regimes is outside of the scope of the paper.

In a long-term perspective, it seems more reasonable to say that  {\it since more than 90\% of bitcoins has already been mined, the effect of the remaining unmined bitcoins on price X must be diminishing.} The more bitcoins are mined, the less effect the last bitcoins will have on $X.$ Additional support for this claim it that there is no danger of running out of bitcoins for transactions. Because of its extreme divisibility, any (practically any) demand can be satisfied with the existing bitcoins.  
 
In the upcoming decade there will be 3 halving events that will reduce the miners' revenue by order of $10.$ Block reward will remain to be the larger part of the miners revenue, provided that there is no decline in the exchange rate X.

\section{Transaction demand}
\label{sec:demand}
In view of arguments that the revenue from the remaining unmined bitcoins will have diminishing effect on the network economy, we can consider a limiting regime where the revenue from transaction fees will be the only source of income of the miners providing the computational resources. This is also in line with the primary function of Bitcoin as an instrument for transacting bitcoins.

 To model the transaction fees, we assume that fees are collected as a $\gamma$--percentage of the transaction value. A  fee depending on the length of the transaction in bytes (as it is implemented now),  or some other complicated fee schedule are also possible and they do not change the logic of the arguments.  It is reasonable to state that the number of demanded transactions, N, of any value is a increasing function of $\gamma.$ Moreover, we make two additional assumptions:

\noindent {\it {\bf H1}: Bitcoin price X is not a factor in determination of the demand for money transfers: $F$ does not depend on $X.$}

This hypothesis seems reasonable. Let’s say you’re a software engineer working in US and you want to buy a new washing mashing for your grand-ma in Brazil. You exchange some of your income in USD for bitcoins, send them to your grand-ma bitcoin wallet and your grand-ma exchange bitcoins for a washing machine in Brazil. As long as, it is done fast, the fluctuations in the exchange rates not likely to affect this transaction. Clearly, this works at any level of the exchange rate X. 

\noindent{\it {\bf H2}: Demand is elastic with respect to parameter $\gamma.$ That is, when $\gamma$ is decreased, the revenue from the fees collected by the network ($F=\gamma VN$) increases, where $V$ is an average transaction value (USD per transaction) .}

The number of transactions (per day) has a maximum bound $N_{max},$ because the number of block (per day) and the size of the block are hard-coded into the network software. It can be estimated using the information about the average size (in bytes) of the transaction. Thus, the revenue is maximized when $\gamma$ reduced to some minimal value $\gamma_{min}$ corresponding to the maximal number of transactions $N_{max}.$

\subsection{Bitcoin network as a whole}

Because miners cannot put pressure on  $\gamma$ by varying the production of hashrate in reasonable limits, the optimal  arrangement is simple.
Consumers would prefer a lower rate $\gamma.$ Miners would like higher revenues and the network designers would like consumers and miners to keep using Bitcoin network. Since the network revenue equals $\gamma VN$ and the demand is elastic, the revenue will be maximized when $\gamma$ is lowest, $\gamma=\gamma_{min}.$ 

By formula \eqref{eq:1} or \eqref{eq:firms}, such $\gamma$ will also maximize the hashrate. The latter property is important for the network designers who would like the hashrate  $H$ to stay above some critical value $H_c,$ below which the network consensus algorithm can be compromised. More precisely, the network relies on the large number of honest computational nodes, and this number presumably correlates with the network hashrate if the  market is dominated by small, independent nodes. Thus, the hashrate can be considered as a proxy to a reliability factor. 


Note that this market equilibrium does not determine the exchange rate X. It can be anything as far as the economics of the transactions is concerned.

\begin{figure}
  \centering
  \includegraphics[width=8cm]{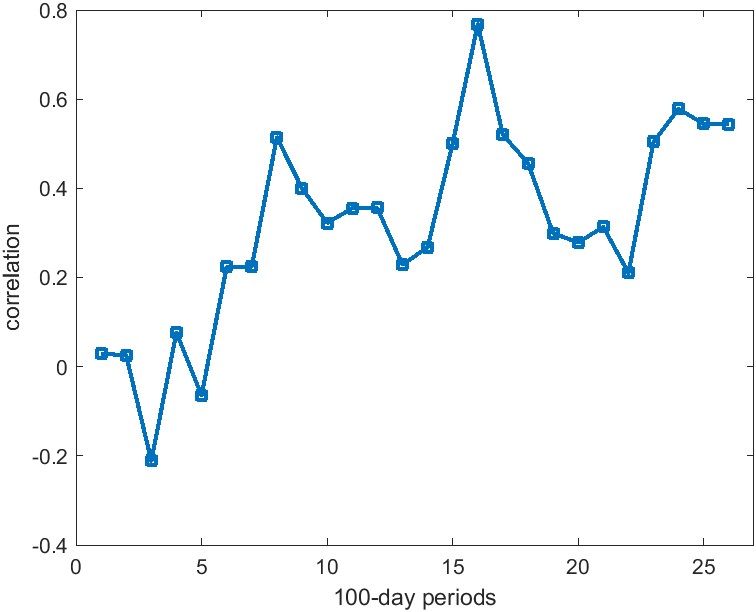}
  \caption{Correlations between daily changes in logarithm of price of Bitcoin and Ethereum over 100-day periods from 08/07/2015 to 10/13/2022. The plot was prepared using data from https://coinmetrics.io/ }
  \label{fig:correlation}
\end{figure}

\section{Conclusions}
At the moment, the  daily value of Bitcoin transaction fees  $F$ is of order $10^{-6}$-- $10^{-4}$ of the value of block reward $BR$.  In one decade $BR$ will be reduced by a factor of 8. If the exchange rate $X$ doesn't increase much by that, as we argued in the previous section, then either
the miners' output will be reduced by the same factor with a modest increase in fees, or the fees must grow enormously. The former outcome seems more realistic, since we saw that the reduction of the hashrate has no effect on the operation of the network.  

Because fluctuations in $X$ are of speculative nature, the reduction in the bitcoin production output and the transition to the regime where the transaction fees will dominate the miners' revenue carries certain risks. 

The reduction of miners' revenue might  negatively affect the speculative price X. For example, miners will be forced to sell bitcoins they have hoarded. This will reduce the revenue even more and undermine the belief in Bitcoin as a mean to store value. As the proportion of such investors (in terms of the capitalization) is huge, their exodus will add a lot of negative pressure on Bitcoin. Thus, it could create a feedback cycle for the revenue and exchange rate X, leading to the crush of Bitcoin price bubble.

There is also risk for Bitcoin is to be outcompeted by other cryptocurrencies. As we mentioned earlier, the reduction of revenue by order 10 will lead to some increase in transaction fees, but even 10 fold increase will not compensate completely for the loss of revenue. Thus, the increase of fees by order 10 is not unreasonable, and since, presently,  transaction fees for Ethereum are about 10 times much as of Bitcoin, 
it is certain possibility that it will outgrow Bitcoin.
 
The competition is already taking place. For example,  the recent improvements in Ethereum technology that now implements a  more energy efficient algorithm of proof-of-stake instead of proof-of-work had direct effect on Bitcoin network that was flooded with former Ethereum miners, who dropped the Bitcoin mining profitability.

We should add here that at this point the competition is not yet reflected in the price dynamics. For example, Figure \ref{fig:correlation} shows the correlation between daily changes in the prices (more precisely changes in the log of the prices) of Bitcoin and Ethereum over 100-day period. Apart from the short period after the introduction of Ethereum, the prices are positively correlated. This shows that neither of the cryptocurrencies is doing better at expense of the other, which might be related to the fact that crypto market is still fresh. There are enough investors' money for all cryptocurrencies and people do not yet differentiate between them.

\end{document}